\documentclass[fullpaper]{jpsj3}

\usepackage{txfonts}
\usepackage{subfigure}
\usepackage{color}

\title{Quantum-classical hybrid algorithm using quantum annealing for multi-objective job shop scheduling}

\author{Kenta Sawamura$^{1,2}$, Kensuke Araki$^{1,2}$, Naoki Maruyama$^{1,2}$, \\Renichiro Haba$^{1}$, and Masayuki Ohzeki$^{1,2,3,4}$}
\inst{%
$^1$Sigma-i Co., Ltd., Tokyo, Japan,
$^2$Graduate School of Information Science, Tohoku University, Sendai, Japan,
$^3$Department of Physics, Institute of Science Tokyo, Tokyo, Japan,
$^4$Semiconductors and Informatics, Kumamoto University, Kumamoto, Japan
}

\abst{Efficient production planning is essential in modern manufacturing to improve performance indicators such as lead time and to reduce reliance on human intuition. While mathematical optimization approaches, formulated as job shop scheduling problems, have been applied to automate this process, solving large-scale production planning problems remains computationally demanding. Moreover, many practical scenarios involve conflicting objectives, making traditional scalarization techniques ineffective in finding diverse and useful Pareto-optimal solutions.
To address these challenges, we developed a quantum-classical hybrid algorithm that decomposes the problem into two subproblems: resource allocation and task scheduling. Resource allocation is formulated as a quadratic unconstrained binary optimization problem and solved using annealing-based methods that efficiently explore complex solutions. Task scheduling is modeled as a mixed-integer linear programming problem and solved using conventional solvers to satisfy detailed scheduling constraints.
We validated the proposed method using benchmark instances based on foundry production scenarios. Experimental results demonstrate that our hybrid approach achieves superior solution quality and computational efficiency compared to traditional monolithic methods. This work offers a promising direction for high-speed, multi-objective scheduling in industrial applications.}

\begin{document}
\maketitle

\section{Introduction}

In modern manufacturing industries, designing efficient production schedules has become increasingly critical in the growing complexity and scale of global supply chains. Such production planning typically aims to minimize metrics such as makespan and lead time and reduce operational costs. These objectives are essential not only for improving productivity but also for ensuring the sustainability of manufacturing operations. Traditionally, skilled personnel manually developed these schedules, relying on their intuition and experience. However, as production scales have increased, this approach has raised concerns about heavier workloads and reliance on individual expertise, which limit its scalability and flexibility. Production planning can be mathematically formulated as a job shop scheduling problem. As a result, increasing interest exists in using mathematical optimization methods to solve job shop scheduling problems systematically and efficiently.

Solving large-scale and complex job shop scheduling problems presents several significant challenges. First, such issues are typically formulated as combinatorial optimization problems. Even with advanced solvers, such as SAT solvers or integer programming solvers, the computational time grows rapidly as the problem size increases, often resulting in exponentially long times.
Furthermore, production planning problems generally involve multiple, and often conflicting, objective functions. These objectives may include minimizing makespan, operational costs, and balancing workload distribution. Due to the trade-offs among these objectives, the problem naturally becomes a multi-objective optimization, where the goal is to identify a diverse and meaningful set of Pareto-optimal solutions.
A common approach to handling multiple objectives is aggregating them into a single objective function using a weighted linear combination. While this method is widely adopted due to its simplicity, it suffers from a well-known limitation: it tends to miss Pareto-optimal solutions located in non-convex regions of the Pareto front. As a result, it may fail to provide a comprehensive solution set necessary for informed and flexible decision-making.

To address these challenges, we have developed a quantum-classical hybrid algorithm for production planning. The core idea of this algorithm is to decompose the problem according to its structural characteristics and delegate each subproblem to an optimization technique best suited for solving it.
In particular, assigning resources, such as furnaces, machines, and personnel, can be formulated as a quadratic unconstrained binary optimization (QUBO) problem. QUBO problems are well-suited for heuristics such as quantum annealing (QA) and simulated annealing (SA). Quantum annealing is a heuristic optimization method specifically designed to solve QUBO problems, which minimizes the following cost function:
\begin{align}
    E(\boldsymbol{q}) = \boldsymbol{q}^\top Q \boldsymbol{q},
\end{align}
where $\boldsymbol{q}$ is a vector of binary variables and $Q$ is a matrix determined by the problem. Practical applications of quantum annealing have been reported in diverse domains, including finance \cite{rosenberg2015solving, orus2019forecasting, venturelli2019reverse}, transportation \cite{neukart2017traffic, hussain2020optimal, inoue2021traffic, shikanai2023traffic}, logistics \cite{feld2019hybrid, ding2021implementation}, warehouse management \cite{ohzeki2019control, haba2022travel}, manufacturing \cite{venturelli2015quantum, yonaga2022quantum}, marketing \cite{nishimura2019item}, materials science \cite{tanaka2023virtual, doi2023exploration}, decoding problems \cite{ide2020maximum, arai2021mean}, and machine learning \cite{neven2012qboost, adachi2015application, benedetti2016estimation, khoshaman2018quantum, o2018nonnegative, amin2018quantum, kumar2018quantum, arai2021teacher, sato2021assessment, urushibata2022comparing, goto2023online, hasegawa2023kernel}. In contrast, SA \cite{kirkpatrick1983optimization} is a classical metaheuristic that searches for the ground state by simulating thermal fluctuations on a computer, rather than exploiting quantum fluctuations. A distinctive advantage of QA is that quantum fluctuations can tunnel through high-energy barriers that are difficult to overcome via thermal fluctuations alone. This feature makes it potentially more effective for problems whose energy landscapes contain steep and narrow valleys, where classical methods are prone to becoming trapped in local minima.

Once resource allocation is determined, the next step is to schedule the processing times of individual tasks. This subproblem can be formulated as a mixed-integer linear programming (MILP) problem. MILP solvers are highly effective in handling complex scheduling constraints and can optimize key performance indicators (KPIs) such as makespan and lead time. 
Combining these two optimization problems, quantum-inspired resource allocation and classical constraint-based scheduling, the proposed hybrid method achieves both flexibility and rigor in solving large-scale production planning problems.

To evaluate the effectiveness of the proposed hybrid approach, we performed experiments using randomly generated benchmark problems inspired by real-world production planning scenarios in the foundry industry. 
In such settings, production planning typically involves a multi-objective optimization problem characterized by the following two KPIs: maximizing the furnace filling ratio and minimizing lead time. 
These two objectives inherently conflict. 
Improving the furnace's filling ratio often requires simultaneous processing tasks with different deadlines. 
This can lead to longer lead times. 
Conversely, prioritizing the reduction of lead time may hinder the efficient use of the furnace.
We adopted the hypervolume indicator to assess the quality of the obtained Pareto-optimal solutions in this multi-objective context. 
This metric quantifies both the convergence and diversity of a Pareto front. 
Using this metric, we compared the performance of our proposed hybrid method with that of a conventional monolithic (non-decomposed) approach. 
The results highlight the hybrid method's superiority in generating a broader and more representative set of trade-off solutions, thereby offering greater flexibility for decision-makers in balancing competing objectives.

The remainder of this paper is as follows: In the next section, we formulate the production planning problem and explain our quantum-classical hybrid algorithm. Next, we construct a set of randomly generated benchmark instances inspired by casting industry production conditions and evaluate the Pareto solution sets obtained by our method. Finally, we summarize our study and discuss future directions toward practical industrial applications.

\section{Methods}

In this section, we describe the proposed quantum-classical hybrid algorithm in detail. We begin by formally defining the manufacturing process scheduling problem that forms the basis of our study.
The scheduling problem in manufacturing can be modeled as a job shop scheduling problem, a well-known class of combinatorial optimization problems. In this formulation, each job consists of multiple tasks associated with various attributes such as processing cost, due date, and required material type. Likewise, each resource representing equipment, such as a furnace or machine, has characteristics such as capacity, setup time for reuse, and processing time. To handle scenarios where a single resource is used repeatedly, we introduced virtual copies of each resource, enabling a more flexible allocation of tasks across time.
Process scheduling aims to determine which resource to assign to each task and at what time, i.e., to define a set of triples (task, resource, start time) that collectively satisfy problem constraints while optimizing performance criteria.
Efficient resource utilization and short lead times are critical objectives in this context. To formally evaluate these aspects, we defined the following KPIs for each combination of task, virtual resource, and processing time:
The first KPI is the resource's filling ratio. This is defined as the ratio of the total processing cost of tasks assigned to a virtual resource to the resource's capacity. A higher filling ratio indicates more cost-effective resource use because minimizing the number of resource activations helps reduce operational overhead.
The second KPI is lead time. Lead time is defined for each task as the absolute difference between its due date and the scheduled processing time. Producing items too early requires additional storage until shipment, while late production risks delivery delays. Hence, minimizing lead time contributes both to operational efficiency and customer satisfaction.
The overarching objective of the scheduling algorithm is to select a set of (task, virtual resource, start time) assignments that jointly maximize the resource filling ratio and minimize lead time, thereby achieving a balanced and optimized production schedule.

\begin{figure}[htb]
\centering
\subfigure[\label{fig:method_non_sep}]{\includegraphics[width=0.45\linewidth]{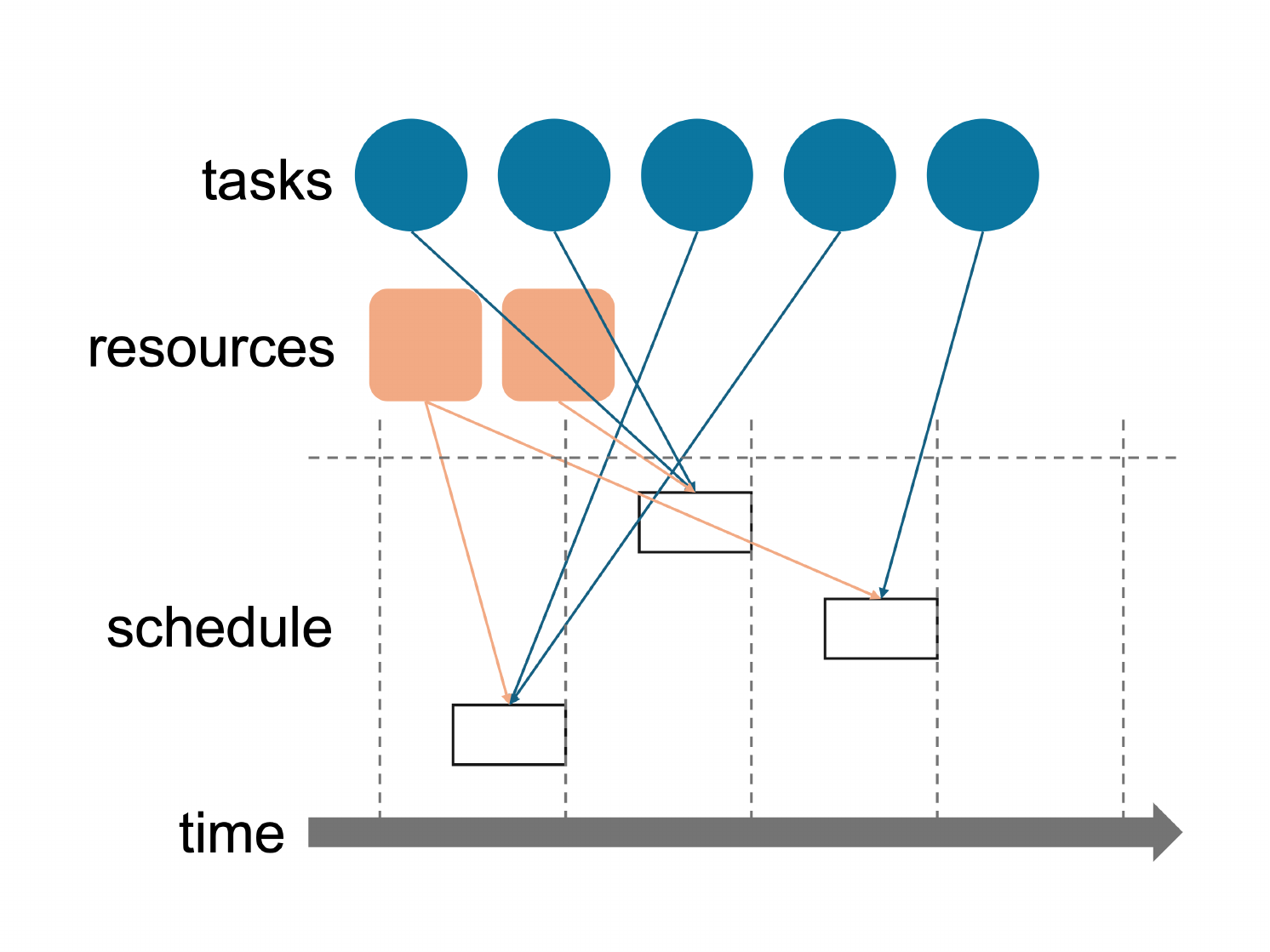}}
\subfigure[\label{fig:method_sep}]{\includegraphics[width=0.45\linewidth]{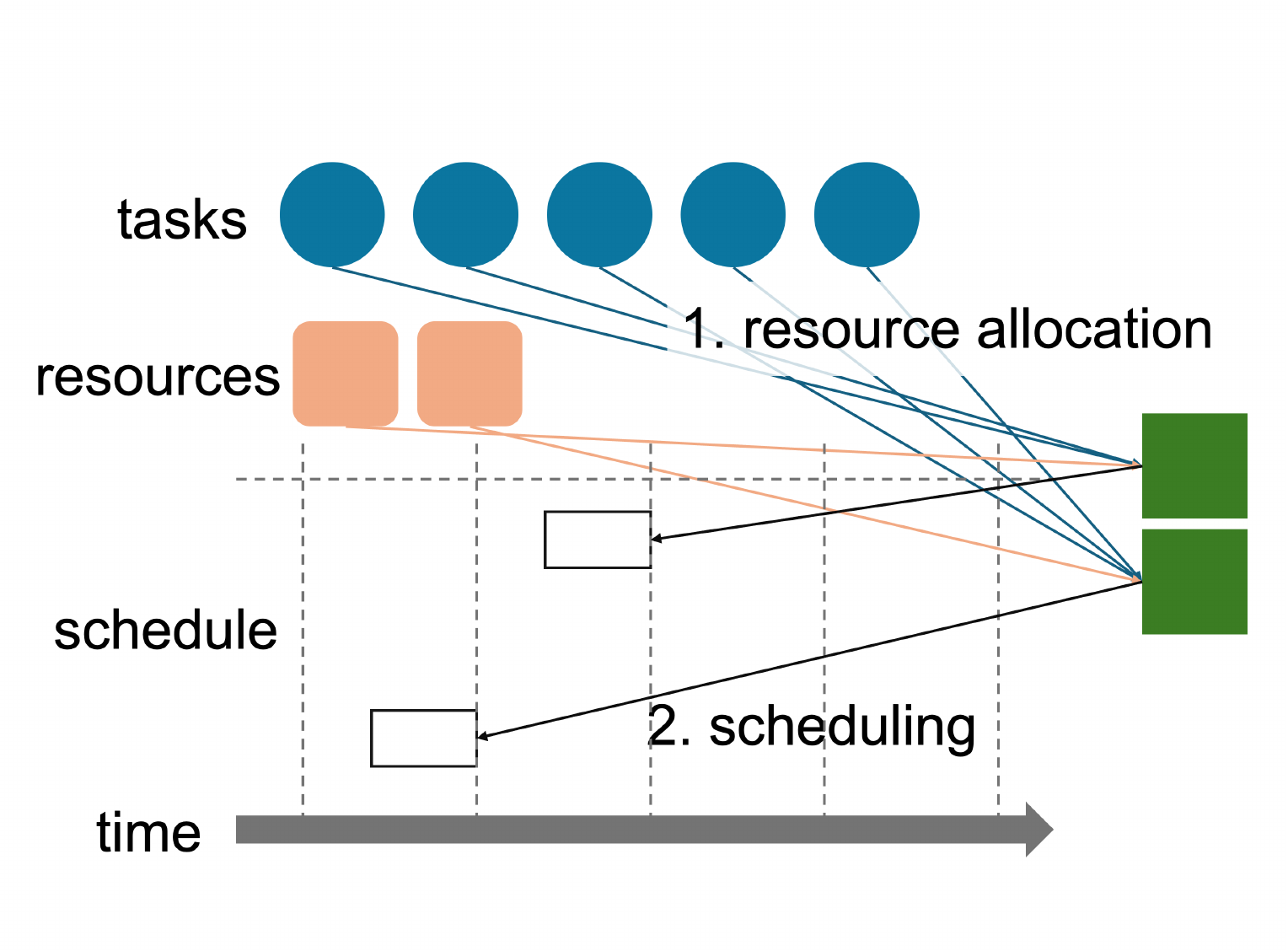}}
\caption{Two modeling approaches for production planning. (a) Non-separation method and (b) separation method.}
\label{fig:method}
\end{figure}

Next, we defined the computational model used to solve the production planning problem as formalized in the previous section. Figure \ref{fig:method_non_sep} shows a straightforward approach to determine, for each task, a virtual resource and a processing time.
However, to efficiently handle the complexity and scale of the problem, we decomposed the production planning problem into two subproblems: a bin packing problem for assigning products to resources, and a scheduling problem for determining the start times of tasks, as illustrated in Fig. \ref{fig:method_sep}.

\subsection{Resource allocation}

We formulated the resource allocation as a combinatorial optimization problem that assigns each task to a feasible resource while minimizing the number of used resources and the dispersion in deadlines. Tasks that can be processed simultaneously on the same resource are grouped into clusters. We assume that the clustering information is provided in advance. A task $i$ belonging to cluster $c$ is denoted as $(c, i)$. For cluster $c$, a virtual resource corresponding to the $l$-th instance of resource $j$ is labeled as $(c, j, l)$. 

We defined decision variables and constants as follows:

\paragraph{Decision Variables}

\begin{itemize}
    \item $y_{cjl} \in \{0,1\}$: Binary variable; equals 1 if processing resource $(c, j, l)$ is used. 
    \item $x_{cijl} \in \{0,1\}$: Binary variable; equals 1 if task $(c, i)$ is assigned to processing resource $(c, j, l)$.
\end{itemize}

\paragraph{Constants}
\begin{itemize}
    \item $C$: Number of task clusters.
    \item $U$: Number of physical resources.
    \item $N_c$: Number of tasks in cluster $c$.
    \item $V_{cj} (\leq N_c)$: Maximum number of virtual processing units for resource $j$ in cluster $c$.
    \item $S_{ci}$: Deadline (target completion date) of task $(c,i)$.
    \item $W_{ci}$: Weight or volume of task $(c,i)$.
    \item $B_j$: Capacity of resource $j$.
\end{itemize}

In the following, we use the notation $[I]:=\{1, \dots, I\}$ to denote the index set for an integer $I$.
We defined the resource allocation optimization problem as follows:

\begin{subequations}
\begin{align}
\min \quad &\sum_{c=1}^{C} \sum_{j=1}^{U} \sum_{l=1}^{V_{cj}} y_{cjl} + a \sum_{c=1}^{C} \sum_{j=1}^{U} \sum_{l=1}^{V_{cj}} \sum_{i=1}^{N_c} \sum_{k=1}^{N_c} (S_{ci}-S_{ck})^2 x_{cijl} x_{ckjl} \label{eq:resource/objective_function/pot number}, \\
\text{s.t.} \quad &\sum_{j=1}^{U} \sum_{l=1}^{V_{cj}} x_{cijl} = 1 \quad \forall c \in [C], i \in [N_c] \label{eq:resource/constraint/one-hot}, \\
&\sum_{i=1}^{N_c} W_{ci} x_{cijl} \leq B_{j} y_{cjl} \quad \forall c \in [C], j \in [U], l \in [V_{cj}] \label{eq:resource/constraint/capacity}, \\
&x_{cijl} \leq y_{cjl} \quad \forall c \in [C], i \in [N_c], j \in [U], l \in [V_{cj}] \label{eq:resource/constraint/xy}, \\
&x_{cijl} \in \{0, 1\} \quad \forall c \in [C], i \in [N_c], j \in [U], l \in [V_{cj}], \\
&y_{cjl} \in \{0, 1\} \quad \forall c \in [C], j \in [U], l \in [V_{cj}]. \\
\end{align}
\end{subequations}
The objective function consists of two terms. The first term of \eqref{eq:resource/objective_function/pot number} minimizes the total number of processing resources used. The second term of \eqref{eq:resource/objective_function/pot number} minimizes the variance in deadlines within the same cluster.
There are three constraints:
Each task is assigned to exactly one resource \eqref{eq:resource/constraint/one-hot}. The total weight of assigned tasks must not exceed the resource capacity \eqref{eq:resource/constraint/capacity}. A task cannot be assigned to an unused resource \eqref{eq:resource/constraint/xy}.

This model is further transformed into a QUBO formulation for compatibility with annealing-based solvers. Each term in the objective and constraints is represented as a quadratic penalty, controlled by corresponding weights $\lambda_{\text{deadline}}, \lambda_{\text{one-hot}}, \lambda_{\text{capacity}}, \lambda_{\text{xy}}$. For inequality constraints such as capacity limits, we relax them into soft constraints using a target filling ratio $\alpha$.

\begin{align}
H &= H_{\text{filling}} + \lambda_{\text{deadline}} H_{\text{deadline}} + \lambda_{\text{one-hot}} H_{\text{one-hot}} + \lambda_{\text{capacity}} H_{\text{capacity}} + \lambda_{\text{xy}} H_{\text{xy}}, \\
H_{\text{filling}} &= \sum_{c=1}^{C} \sum_{j=1}^{U} \sum_{l=1}^{V_{cj}} y_{cjl}, \\
H_{\text{deadline}} &= \sum_{c=1}^{C} \sum_{j=1}^{U} \sum_{l=1}^{V_{cj}} \sum_{i=1}^{N_c} \sum_{k=1}^{N_c} \left(S_{ci}-S_{ck}\right)^2 x_{cijl} x_{ckjl}, \\
H_{\text{one-hot}} &= \sum_{c=1}^{C} \sum_{i=1}^{N_c} \left(\sum_{j=1}^{U} \sum_{l=1}^{V_{cj}} x_{cijl} - 1\right)^2, \\
H_{\text{capacity}} &= \sum_{c=1}^{C} \sum_{j=1}^{U} \sum_{l=1}^{V_{cj}} \left(\sum_{i=1}^{N_c} w_{ci} x_{cijl} - \alpha B_{j} y_{cjl}\right)^2, \\
H_{\text{xy}} &= \sum_{c=1}^{C} \sum_{i=1}^{N_c} \sum_{j=1}^{U} \sum_{l=1}^{V_{cj}}  \left(y_{cjl} - x_{cijl} - \frac{1}{2}\right)^2.
\end{align}

\subsection{Scheduling}

Given the resource allocation results, each assigned resource $j$ and its $l$-th processing can be represented as $(j,l)$.  
The scheduling stage determines the optimal completion time for each task, minimizing lead time while respecting machine exclusivity constraints.

\paragraph{Decision Variables}
\begin{itemize}
    \item $z_{cjl} \in \mathbb{R}$: Completion time of the $l$-th processing on resource $j$ for cluster $c$.
    \item $t_{cjl} \in \mathbb{R}$: Auxiliary variable representing the absolute deviation between the actual completion time and the target completion deadline.
    \item $s_{cjl_1l_2} \in \{0,1\}$: Binary variable; equals 1 if task $(c,j,l_1)$ is scheduled before task $(c,j,l_2)$ on the same resource.
\end{itemize}

\paragraph{Constants}
\begin{itemize}
    \item $U$: Number of physical resources.
    \item $V_j$: Number of processing tasks assigned to resource $j$.
    \item $F_{cjl}$: Target completion deadline for task $(c,j,l)$.
    \item $R_{cjl}$: Processing time required for task $(c,j,l)$.
    \item $T_j$: Setup time required for resource $j$ between consecutive tasks.
    \item $M$: A sufficiently large constant for linearizing disjunctive constraints (Big-$M$ method).
\end{itemize}
The scheduling problem is formulated as:
\begin{subequations}
    \begin{align}
        \min \quad &\sum_{c=1}^{C} \sum_{j=1}^{U} \sum_{l=1}^{V_j} t_{cjl} \label{eq:scheduling/objective_function/lead-time}, \\
        \text{s.t.} \quad
        &z_{cjl_1} + T_{j} \leq z_{cjl_2} - R_{cjl_2} + M \cdot s_{cjl_1l_2}, \\
        &z_{cjl_2} + T_{j} \leq z_{cjl_1} - R_{cjl_1} + M \cdot (1-s_{cjl_1l_2}), \notag \\
        &\hspace{2cm} \forall c, \forall j, \; l_1 \neq l_2 \in [V_j] \label{eq:scheduling/constraint/exclusion}, \\
        &-t_{cjl} \leq z_{cjl} - F_{cjl} \leq t_{cjl}, \quad \forall c, j, l, \\
        &z_{cjl}, t_{cjl} \in \mathbb{R}, \quad \forall c, j, l, \\
        &s_{cjl_1l_2} \in \{0, 1\}, \quad \forall c, j, \; l_1 \neq l_2.
    \end{align}
\end{subequations}
The objective function \eqref{eq:scheduling/objective_function/lead-time} minimizes the sum of absolute deviations between actual completion times and their respective target deadlines, thereby reducing overall lead time.  
Constraint \eqref{eq:scheduling/constraint/exclusion} enforces resource exclusivity, ensuring that no two different processing $(c,j,l_1)$ and $(c,j,l_2)$ overlap in time.  
The big-$M$ formulation allows modeling disjunctive sequencing decisions using binary variables $s_{cjl_1l_2}$.  
Finally, the auxiliary variables $t_{cjl}$ capture deviations in a linear form, enabling the minimization of absolute values within an MILP framework.

To solve these two subproblems efficiently, we adopted different optimization techniques tailored to their respective characteristics. For the resource allocation problem, we used annealing-based methods, such as QA or SA, which are well-suited for rapidly sampling high-quality solutions in high-dimensional, nonlinear search spaces. For the scheduling problem, we used an MILP solver, which offers precise control over constraint satisfaction and optimization of temporal objectives such as makespan and lead time.
This hybrid modeling strategy enables both computational efficiency and high solution quality, effectively leveraging the strengths of distinct optimization paradigms within a unified framework.

\section{Results}
This section comprehensively evaluates the proposed hybrid approach using randomly generated instances inspired by real-world casting industry scenarios. In such production planning, limited melting furnace capacity must be allocated among multiple tasks while achieving timely delivery and high production efficiency. Filling ratio and lead time are often conflicting KPIs, making it difficult to obtain a single optimal solution. In this study, we formulated the problem as a multi-objective optimization problem that optimizes both KPIs and compared the proposed method against a conventional baseline.

As the baseline, we used the non-decomposed method, which solves the problem in a single step. In contrast, the proposed method decomposes the problem into resource allocation and scheduling stpdf, applying solvers suited to each subproblem in a hybrid configuration. 

As a benchmark, we generated random instances inspired by process planning in the casting industry. Each instance consists of a single melting furnace (capacity = 20, setup time = 0) and a set of tasks representing product castings. We prepared 10 instances for each problem size with $\{6, 8, 10, 12\}$. All tasks are assumed to belong to a single cluster $c$, and each task is labeled as $(c, i)$, where $i$ indicates the $i$-th task within cluster $c$. For each task $(c, i)$, the weight $w_{(c, i)}$ was independently sampled from the discrete uniform distribution $U\{1, 10\}$, and the due date $d_{(c, i)}$ from $U\{3, 30\}$. These instances simulate diverse order patterns observed in actual manufacturing environments. The conditions of the solvers are listed in Table \ref{table:experiment_condition}.

\begin{figure}[htb]
\centering
\subfigure[\label{fig:parato}]{\includegraphics[width=0.45\linewidth]{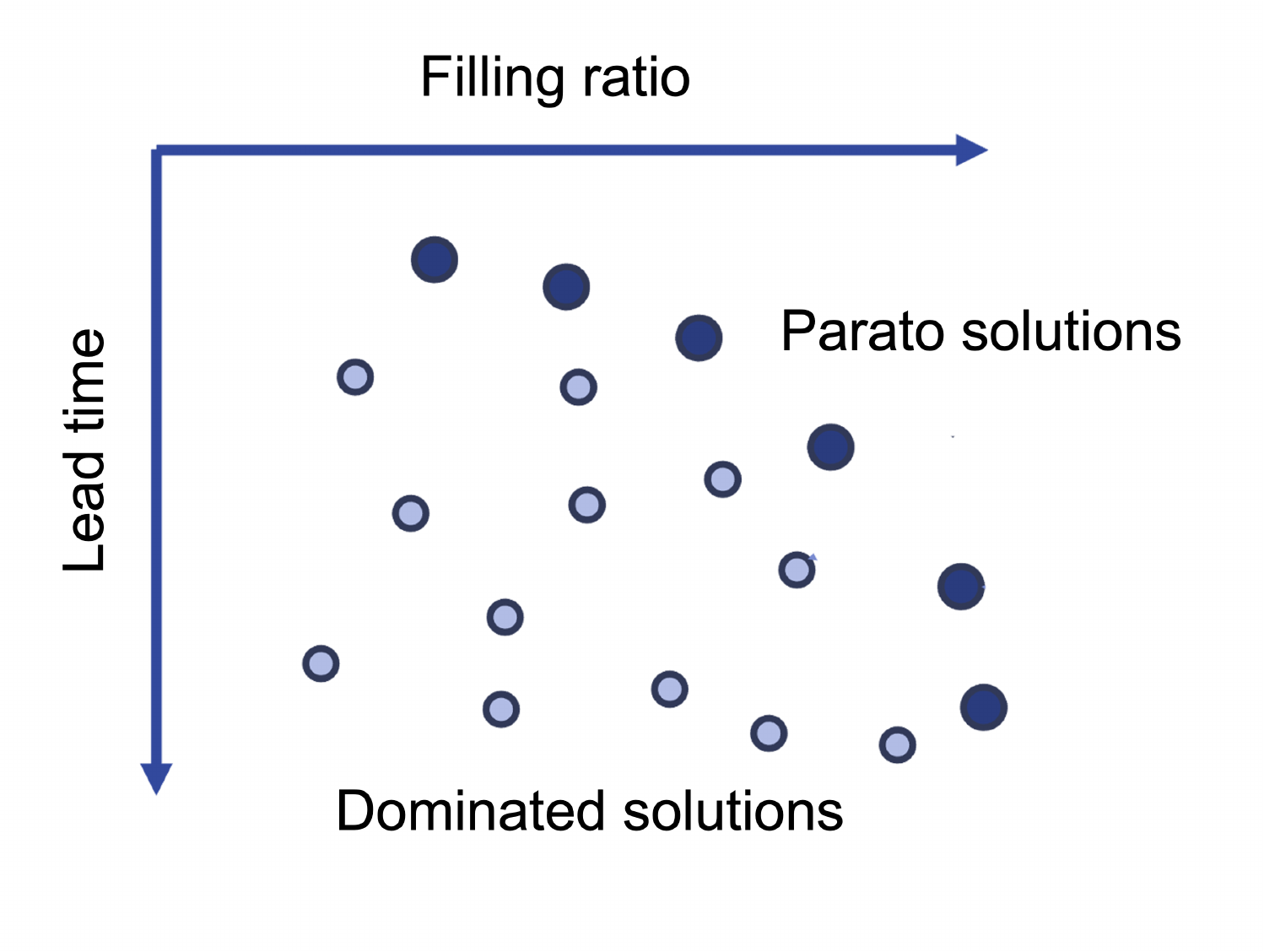}}
\subfigure[\label{fig:hypervolume}]{\includegraphics[width=0.45\linewidth]{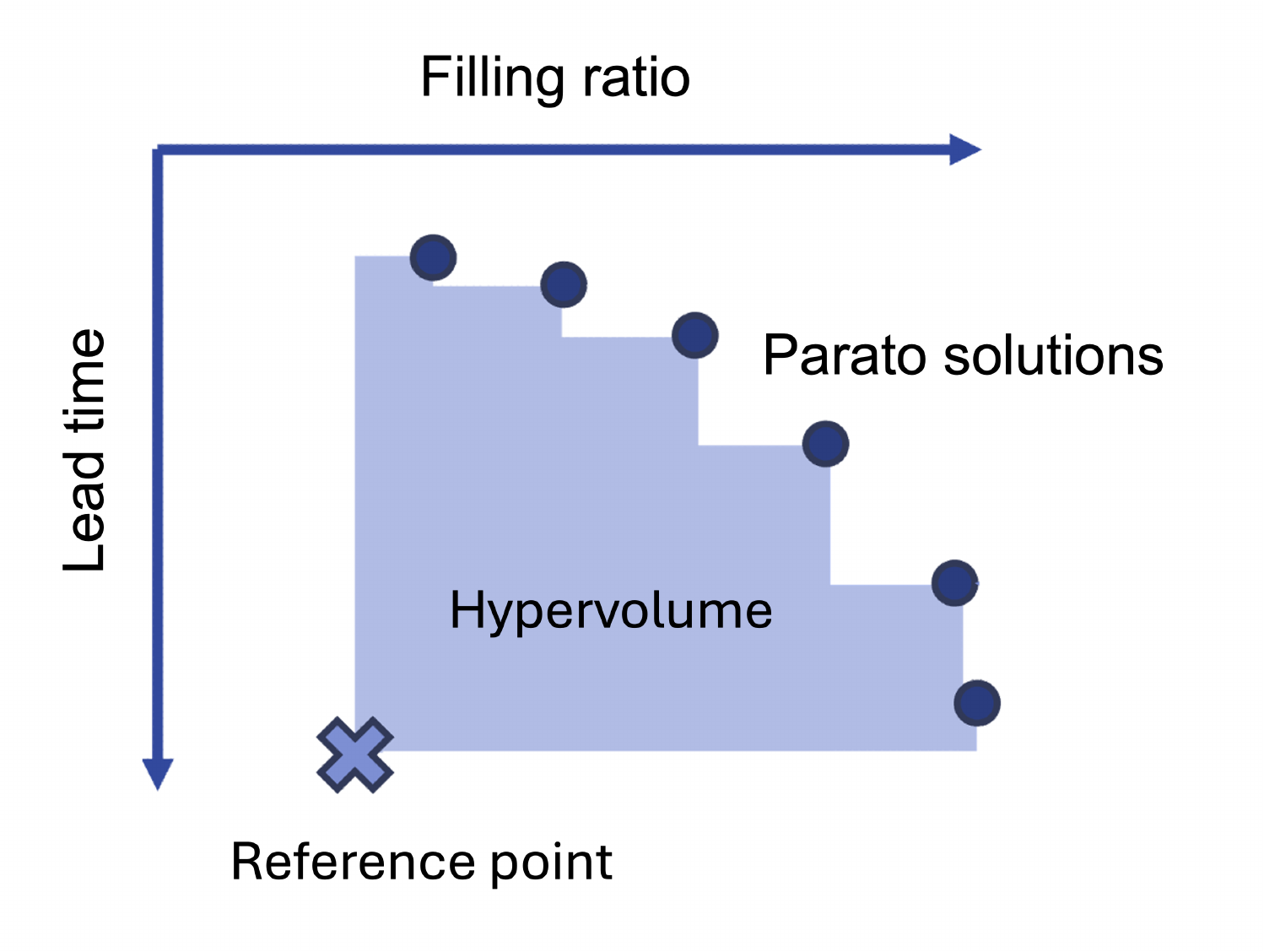}}
\caption{Conceptual diagram of Pareto optimal set and hypervolume metric. (a) Parato solutions and (b) hypervolume.}
\label{fig:parato_and_hypervolume}
\end{figure}

\begin{table}[h]
  \centering
  \caption{The conditions of the solvers. (a): ortools.sat.python-cp\_model, (b): dwave-neal, (c): ortools.linear\_solver-pywraplp, (d): dwave-Advantage\_system6.4, (e): embedding\_composite}
  \label{table:experiment_condition}
  \begin{tabular}{ccccc} \hline
     method & solver & sampler & num\_reads & timeout(sec) \\ \hline
     non-separation & (a) & - & - & 0.1 \\
     separation (SA) & (b), (c) & - & 1000 & 0.1 \\
     separation (QA) & (d), (c) & (e) & 1000 & 0.1 \\ \hline
  \end{tabular}
\end{table}
To evaluate solution quality in the multi-objective setting, we adopted the hypervolume metric, which measures the volume of the objective space dominated by the obtained Pareto-optimal set. Figure \ref{fig:parato} illustrates a Pareto-optimal set, which contains solutions not dominated by any other solution across all objectives and offers a valuable set of trade-off options for decision-makers. Figure \ref{fig:hypervolume} shows that larger hypervolume values indicate higher quality and broader diversity of solutions.

\begin{figure}[htb]
\centering
\subfigure[\label{fig:SA_non_sep}]{\includegraphics[width=0.45\linewidth]{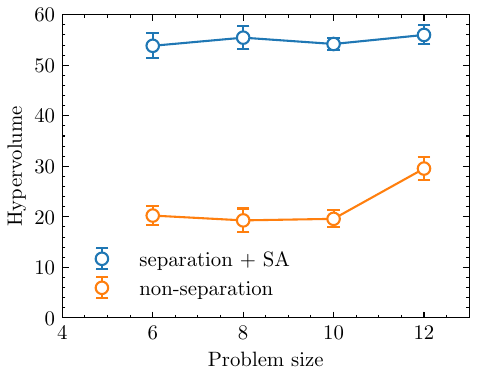}}
\subfigure[\label{fig:QA_non_sep}]{\includegraphics[width=0.45\linewidth]{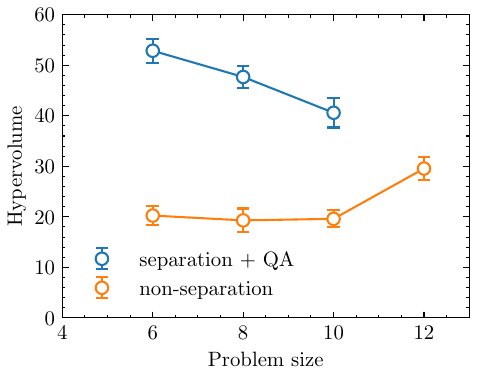}}
\caption{Comparison of hypervolume between the separation method and the non-separation method. Horizontal axis is the number of orders (problem size) and vertical axis is hypervolume. (a) Separation method using SA and (b) separation method using QA.}
\label{fig:SAQA_non_sep}
\end{figure}

\begin{figure}[ht]
\centering
\includegraphics[width=0.5\linewidth]{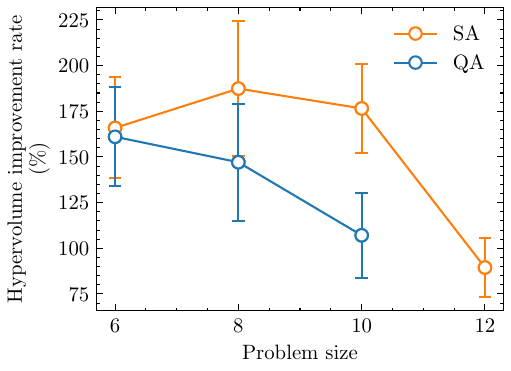}
\caption{Improvement rate of the separation method and the non-separation method. Horizontal axis is the number of orders (problem size) and vertical axis is the improvement rate of hypervolume.}
\label{fig:improvement_rate_SAQA_non_sep}
\end{figure}

Figure \ref{fig:SAQA_non_sep} showed substantial performance gains for the proposed method. As shown in Fig. \ref{fig:improvement_rate_SAQA_non_sep}, with SA, the hypervolume improved by up to 180\% compared to the non-decomposed method, while with QA, it improved by up to 160\%. The non-decomposed method produced solutions biased toward a single KPI, either filling ratio or lead time, resulting in narrower Pareto coverage. In contrast, the hybrid approach produced solutions that balanced both KPIs, allowing the Pareto set to dominate a wider range of inferior solutions.

We further examined robustness by randomly perturbing resource allocations and observing performance changes. The procedure was as follows: (1) compute the initial resource allocation and schedule, (2) randomly select two tasks and swap them, adopting the swap only if capacity constraints were satisfied, (3) recompute the schedule for the modified allocation, and (4) compare the hypervolume before and after the swap. Smaller changes in hypervolume indicate structurally stable, robust solutions.

\begin{figure}[htb]
\subfigure[]{\includegraphics[width=0.45\linewidth]{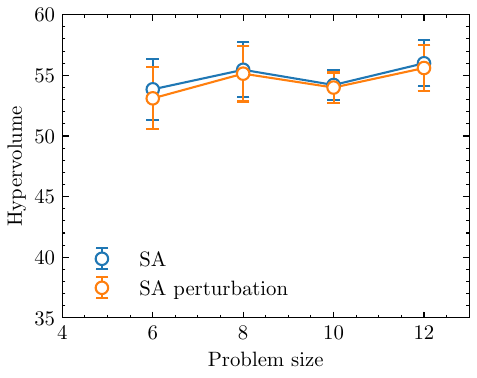}}
\subfigure[]{\includegraphics[width=0.45\linewidth]{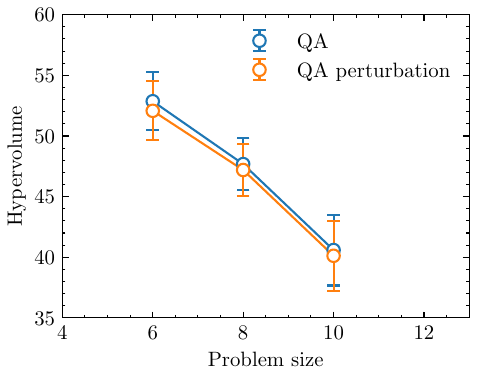}}
\caption{Comparison of robustness in the separation method. Horizontal axis is the number of orders (problem size) and vertical axis is hypervolume. Preturbation is the result of recalculating the schedule by replacing part of the resource allocation. (a) Separation method using SA and (b) separation method using QA.}
\label{fig:robust}
\end{figure}

\begin{figure}[ht]
\centering
\includegraphics[width=0.5\linewidth]{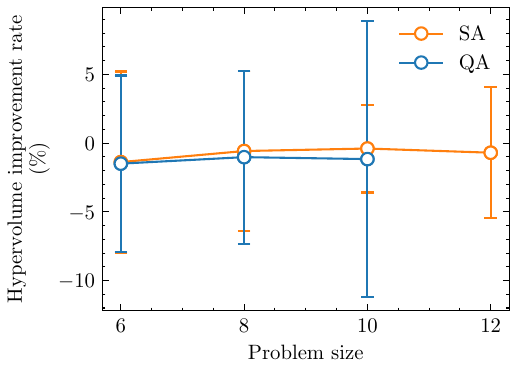}
\caption{Improvement rate in robustness. Horizontal axis is the number of orders (problem size) and vertical axis is the improvement rate in hypervolume (negative values indicate deterioration).}
\label{fig:improvement_rate_robust}
\end{figure}

Figure \ref{fig:robust} showed that hypervolume decreased after swaps in all cases, suggesting that the original allocations corresponded to local optima. Moreover, as shown in Fig. \ref{fig:improvement_rate_robust},  QA-based solutions exhibited slightly greater performance degradation than SA-based ones, possibly reflecting a stronger dependence on specific allocation patterns inherent to QA solutions. These findings demonstrate the proposed method's superior performance over conventional approaches and provide valuable insights into the structural properties and stability of the solutions.

\section{Discussion}
In this study, we proposed and developed a quantum–classical hybrid algorithm for solving process planning problems. To evaluate its effectiveness, we conducted experiments using randomly generated test cases that simulate practical casting operations, comparing the results against a conventional non-decomposed method. The results confirmed that our method can create broader, higher-quality Pareto solution sets. In particular, the proposed approach can balance multiple objectives (e.g., filling ratio and lead time) more effectively, enabling practitioners to select production schedules simultaneously satisfying multiple KPIs. Although the experiments in this study focused on a production planning problem in the casting industry, the proposed method broadly applies to scheduling and resource allocation problems in other manufacturing domains. By appropriately adapting the problem formulation and constraints, the hybrid algorithm can be extended to diverse industrial contexts where multiple conflicting objectives must be optimized simultaneously. This capability has significant implications for many manufacturing domains where complex trade-offs among objectives are inherent.

The decomposition framework plays a central role in the algorithm performance: the bin-packing subproblem (resource allocation) is solved using QA or SA for fast and diverse sampling, followed by solving the scheduling subproblem via a classical MILP solver to ensure precision and constraint satisfaction. This exploratory breadth and computational accuracy combination is particularly well-suited for real-world problems characterized by multiple objectives and intricate constraints.

Future work will focus on scaling the method to more realistic, larger problem instances. Increasing the problem size in QA often leads to a higher frequency of constraint-violating solutions, making it challenging to extract a valid solution set. Addressing this will require improved parameter tuning strategies and the design of algorithms that inherently respect hard constraints. Furthermore, in actual factory environments, disturbances such as sudden order changes or shifts in resource availability can occur after planning. In such scenarios, producing robust schedules with stable KPIs is crucial despite these disruptions.

Although prior studies have suggested that QA is inherently more likely to yield robust solutions, our results showed the opposite trend: SA produced more robust outcomes. This counterintuitive finding suggests that the interaction between the annealing method used for resource allocation and the hypervolume metric may play a role. However, the precise relationship between allocation robustness and the resulting hypervolume remains unclear. Further experimental studies and theoretical analyses are necessary to reveal this relationship and fully understand the robustness behavior of quantum–classical hybrid process planning.

\begin{acknowledgment}
This paper is based on results obtained from a project, JPNP23003, commissioned by the New Energy and Industrial Technology Development Organization (NEDO).
\end{acknowledgment}

\section*{Author contributions}
K.S. conceived of the presented idea, developed the methodology, and performed the experiments. K.A. managed the project. N.M. supervised it. R.H. helped to supervise it. M.O. reviewed the draft. All authors discussed the results and contributed to the final manuscript.

\bibliographystyle{jpsj}
\bibliography{reference}

\end{document}